# How Digital Transformation Impacts Corporate Green Innovation?

Chen Hanqin

**Abstract**：Digitalization is a defining feature of our time, and green innovation has become one of the necessary avenues for firms to achieve sustainable development. Using financial statements and annual report data for China's A-share listed firms from 2010–2019, this paper constructs a firm-level digital transformation indicator and examines the impact of digital transformation on green innovation and its mechanisms. The results show that digital transformation promotes firms' green-innovation output, with a diminishing marginal impact over time. Mechanism tests indicate that digital transformation boosts green-innovation output by increasing R&D investment and strengthening environmental management. Heterogeneity analysis shows that the promotion effect is more pronounced for small and medium-sized firms and for firms in technology-intensive industries. To improve the green-innovation incentives of digital transformation, firms should formulate long-term strategies and continuously strengthen policy regulation and incentives.

## I. Introduction

The rise of digitalization is a salient feature of economic and social development since the beginning of the twenty-first century. Since the Eighteenth National Congress of the Communist Party of China, the Central Committee has attached great importance to developing the digital economy and has elevated it to a national strategy. The report to the Twentieth National Congress further explicitly proposed to accelerate the building of a strong manufacturing country, a strong quality country, a strong aerospace country, a strong transportation country, a strong cyber power, and a Digital China. At present, a surging wave of informatization is sweeping the globe, and countries worldwide regard the advancement of economic digitalization as an important driving force for achieving innovative development. The digital economy is the core of the economic transformation of China, and the degree of enterprise digital transformation is the foundation for achieving this objective. In the era of digitalization, an increasing number of enterprises have come to realize the importance of digital transformation for their business operations and sustainable development, and green innovation has become one of the necessary pathways for enterprises to achieve sustainable development.

Although over the past decade the digital economy of China has flourished and the scale of the industry has continued to grow rapidly, ranking second in the world for many years, many small and medium-sized enterprises in China are still at the initial stage of digital transformation, and only a small share of enterprises have reached the stage of deep application. The China Digital Economy Development Report (2022) shows that from 2012 to 2021 the scale of the digital economy of China increased from 11 trillion yuan to 45.5 trillion yuan, and its share of gross domestic product rose from 21.6 percent to 39.8 percent. The data indicate that more than ninety percent of enterprises remain at the normative stage of digital transformation, and fewer than ten percent of enterprises have reached the domain level. Against the backdrop that the overall digital transformation of enterprises in China is still at an initial stage, does enterprise digital transformation have a significant effect on green innovation. If there is a positive effect, through which channels and mechanisms does it promote green innovation output. Furthermore, which types of enterprises are more adept at achieving growth in green innovation output through digital transformation.

As the main driving forces of the new round of scientific and technological revolution, digitalization and informatization have become important subjects of academic research. Existing studies mainly focus on the effect of information technology on enterprise innovation. In the early stage of informatization, information technology creates value at the intermediate stage of the production process by raising innovation productivity, and it plays a key role in promoting breakthrough innovation through research and development and other intangible factors such as skills and knowledge (Kleis et al., 2012). Informatization construction can also significantly improve the quality of patents of enterprises and promote the growth of invention patents, rather than only driving short-cycle practical utility models and design patents, and it raises the output efficiency of invention patents of enterprises (Li Lei et al., 2022). As an advanced stage of the development of informatization, digitalization places greater emphasis on the comprehensive and in-depth application of information technology to create value. Enterprise digital transformation can improve environmental impact by promoting green technological innovation, enhancing environmental information disclosure, and strengthening governance (Qiong Xu et al., 2021). Digital transformation also promotes enterprise innovation through mechanisms such as realizing open and networked innovation, leading organizational and managerial innovation, and raising the level of human capital within enterprises (An Tongliang, 2022).

Although digital transformation has attracted wide attention in the field of enterprise innovation, research on its effect on green innovation remains relatively limited. In recent years, with the continuous enhancement of environmental awareness, green innovation has become a new pathway for enterprises to achieve green, efficient, and sustainable development (Chen Zewen and Chen Dan, 2019). As a comprehensive transformation, digital transformation encompasses a variety of emerging technologies and is bound to have a profound effect on the green innovation of enterprises. For example, in production and operations, digital technologies can help enterprises realize efficient, precise, and sustainable utilization of resources, thereby reducing negative environmental impact. Based on this, and taking the strategic goal of building a Digital China as an opportunity, this paper examines how digital transformation affects the green innovation of enterprises, with the aim of providing ideas and references for enterprises on the road toward sustainable development. This paper measures the degree of enterprise digital transformation by the frequency of digital-transformation-related keywords in enterprise annual reports, and measures enterprise green innovation output activity by the number of green patent applications, in order to study the effect of digital transformation on green innovation output of enterprises. Building on existing research, the innovations of this paper are mainly reflected in the following two aspects. First, with respect to empirical methodology, because the number of green patent applications is a non-negative integer and the dependent variable is a count variable, prior studies have often employed log(y+1) together with ordinary least squares for estimation; however, this common log-linear approach yields estimates that lack meaningful economic interpretation and exhibit inherent bias. This paper adopts a Poisson pseudo maximum likelihood estimator with multi-dimensional fixed effects drawn from count models, which can minimize bias caused by omitted variables to the greatest extent possible. Second, from the perspective of analyzing internal mechanisms, this paper devotes part of the analysis to enterprise environmental management. Digital transformation can standardize internal processes of enterprises and thereby, through digital transformation, form better incentives for green innovation. However, existing research has relatively seldom examined green innovation output from the angle of enterprise environmental management. Therefore, from this perspective, this paper explores how digital transformation helps enterprises realize better incentives for green innovation and provides new implications for relevant policy and practice.

## II. Theoretical Analysis and Research Hypotheses

Environmental management and R&D-investment channels are closely interconnected. First, the environmental-management channel of firms' digital transformation can support and propel the R&D-investment channel. Through digital environmental-management practices, firms can achieve efficient utilization of environmental resources and reduce pollution, thereby enhancing the sustainability and environmental friendliness of their R&D-investment activities. Second, the R&D-investment channel of digital transformation can in turn foster innovation and optimization in environmental management. By leveraging digital tools, firms can conduct R&D and innovation more efficiently, thus accelerating innovation and improvement within the environmental-management channel. Drawing on prior studies, this paper selects the environmental-management channel and the R&D-investment channel as two key mechanisms through which digital transformation operates.

2.1 Environmental-Management Channel

Environmental-management certification is one of the metrics for firms' innovation activities, because it reflects both the firm's emphasis on environmental protection and its innovative capability in environmental management. As awareness has grown regarding the environmental impacts of industrial activities, various standards have been formulated to guide firms and to incorporate environmental-management systems into corporate assessments. Among these, the ISO 14001 Environmental Management System—formally introduced in September 1996—has been the most widely adopted worldwide. Because ISO 14001 requires firms to set internal environmental standards, targets, and performance indicators, whether a firm has obtained this certification can be used to gauge its level of environmental-management innovation (Peiyan Zhou et al., 2022). At the national level, participation in ISO 14001 is an important predictor of a country's environmental patenting and serves as a standard indicator of innovative activity (S. Lim et al., 2014).

Innovation constitutes a firm's core competitiveness. Therefore, firms holding ISO 14001 certification possess more potential competitive advantages, including higher internal efficiency, differentiation advantages, responsiveness to stakeholder demands, improved industry positioning, and financial savings (Murillo-Luna et al., 2008). At the same time, managerial decision-making plays an important role in strengthening competitiveness: managers with strong environmental awareness proactively pursue green-innovation R&D and cleaner production, actively develop environmentally friendly products, and thereby enhance the firm's market competitiveness (Li Hui et al., 2022).

Based on the foregoing analysis of the environmental-management channel, we propose the following hypothesis:

H1: Digital transformation promotes firms' green-innovation output by enhancing environmental-management capability.

## 2.2 R&D-Investment Channel

R&D is the source of firms' innovation and is closely related to innovative outcomes. First, digital transformation can directly improve process-innovation performance and product-innovation performance in manufacturing enterprises. On the one hand, digital transformation drives modularization, automation, and intelligentization in production, thereby raising efficiency and improving production methods; on the other hand, it facilitates information flows in product-design departments, broadening and deepening information sources and stimulating firms' innovation capabilities in new-product development (Shuhao Liang et al., 2022). Second, digital transformation can also increase firms' R&D investment. By lowering internal and external control costs, firms can allocate freed-up resources to R&D and innovation—for example, to pollution-control equipment—thus raising the level of green innovation (Shujun Sun et al., 2022). This helps firms better fulfill social responsibility and enhance their image in environmental protection and sustainable development, enabling them to meet market demand more effectively and to drive sustained innovation and growth.

In sum, digital transformation exerts a positive, facilitating effect on firms' R&D and innovation activities. By boosting production-process efficiency and information fluidity and by increasing R&D investment, digital transformation creates favorable conditions for innovation and is expected to provide strong support for firms to maintain competitiveness and achieve sustainable development in highly competitive markets.

Based on the foregoing analysis of the R&D-investment channel, we propose the following hypothesis:

H2: Digital transformation promotes firms' green-innovation output by increasing R&D investment.

# III. Research Design

## 3.1 Data Description

This study uses green patent data for A-share listed enterprises in the Shanghai and Shenzhen stock exchanges in China from 2010 to 2019, together with enterprise-level economic data for the corresponding enterprises. For the above database of listed enterprises, the following sample screening and processing are conducted: (i) enterprises in the financial industry are removed, and only enterprises belonging to the real economy are retained; (ii) samples that are subject to special treatment in stock trading, that is, stock codes containing ST or *ST, as well as enterprises with missing key financial indicators, are removed; and (iii) in order to eliminate the influence of extreme values, all continuous variables are winsorized at the 1 percent level. The final sample includes 1,512 enterprises, with 15,120 enterprise-year observations.

The green innovation output capability of enterprises is measured as follows. Green patent data for listed enterprises come from annual reports of listed enterprises, corporate social responsibility reports of listed enterprises, official websites of listed enterprises, the National Bureau of Statistics, the China National Intellectual Property Administration, and the World Intellectual Property Organization, and green patents of enterprises are identified with the aid of the environmental International Patent Classification index list released by the World Intellectual Property Organization in 2010. The data fields used in this paper include the total number of green patent applications, as well as counts of invention patent applications and utility model patent applications classified by invention type.

The degree of digital transformation of enterprises is obtained from the CSMAR "Research Database on Digital Transformation of Chinese Listed Enterprises," and the enterprise digitalization level is measured by the frequency of keywords related to digital transformation in annual reports.

Enterprise economic data for listed enterprises all come from the CSMAR database. Drawing on prior literature, a series of control variables are selected, mainly including enterprise size, enterprise age, board size, state-ownership dummy, asset-liability ratio, capital intensity, Tobin Q, ISO 14001 certification dummy, year dummies, and industry dummies. The specific variable settings are reported in Table 1.

Table 1. Variable definitions

| Type | Name | Symbol | Measurement |
|---|---|---|---|
| Dependent variables | Green patent applications in year t | apply0 | Total number of green patent applications of the enterprise in year t |
| | Green patent applications in year t+1 | apply1 | Total number of green patent applications of the enterprise in year t+1 |
| | Green patent applications in year t+2 | apply2 | Total number of green patent applications of the enterprise in year t+2 |
| | Green invention patent applications in year t | invention0 | Total number of green invention patent applications of the enterprise in year t |
| | Green invention patent applications in year t+1 | invention1 | Total number of green invention patent applications of the enterprise in year t+1 |
| | Green invention patent applications in year t+2 | invention2 | Total number of green invention patent applications of the enterprise in year t+2 |
| | Green utility model patent applications in year t | utility0 | Total number of green utility model patent applications of the enterprise in year t |
| | Green utility model patent applications in year t+1 | utility1 | Total number of green utility model patent applications of the enterprise in year t+1 |
| | Green utility model patent applications in year t+2 | utility2 | Total number of green utility model patent applications of the enterprise in year t+2 |
| Core explanatory variables | Enterprise digital transformation | digital | ln(total frequency of all digitalization-related keywords in the enterprise annual report in year t + 1) |
| | Artificial intelligence technology | ai | ln(frequency of artificial intelligence terms in the report + 1) |
| | Blockchain technology | bc | ln(frequency of blockchain terms in the report + 1) |
| | Cloud computing technology | cd | ln(frequency of cloud computing terms in the report + 1) |
| | Big data technology | bd | ln(frequency of big data terms in the report + 1) |
| | Digital technology application | dt | ln(frequency of digital application terms in the report + 1) |
| Control | Enterprise size | size | ln(total assets at year-end) |

| variables | | | |
|---|---|---|---|
| Enterprise age | age | Current year minus the year of listing of the enterprise | |
| Board size | bm | Total number of directors on the board | |
| State-owned enterprise | state | Equals 1 if a state-owned enterprise, otherwise 0 | |
| Leverage | leverage | Total liabilities of the enterprise divided by total assets of the enterprise | |
| Capital intensity | intensity | Total assets of the enterprise divided by operating revenue | |
| Tobin Q | tobinq | Market value of the enterprise divided by replacement cost of capital | |
| ISO 14001 certification | ISO14001 | Equals 1 if ISO 14001 audit has been passed, otherwise 0 | |
| Year dummies | year | Year dummies for 2010–2019 | |
| Industry dummies | industry | Classified according to the 2012 China Securities Regulatory Commission industry classification | |

3.2 Model specification and econometric method

In order to examine the effect of digital transformation on the green innovation output capability of enterprises, a multiple linear regression model is constructed as shown in Model (1).

$$\text{Innovation}_{ijt} = \alpha + \beta_1 \text{digital}_{jt} + X'\Gamma + \omega_t + \eta_i + \varepsilon_{it} \qquad (1)$$

（1）Dependent variable

Green innovation output capability of enterprises (Innovation). This paper chooses green patent application data, which are less affected by external factors and are more stable, to measure the green innovation output capability of enterprises. Under the patent system of China, patents are divided into invention patents, utility model patents, and design patents. Among them, invention patents have the highest innovation value, while utility model and design patents have lower innovation value, because invention patents refer to novel and thorough improvements to products or processes, whereas utility models are innovations in technical solutions considered from the perspective of technical effects and functions of products. In other words, invention and utility model patents possess the characteristics of novelty, inventiveness, and utility, whereas design patents do not involve the technical performance of the product itself. Therefore, after comprehensive consideration, this paper selects the numbers of invention patent applications and utility model patent applications for discussion. $\text{Innovation}_{ijt}$ denotes the green innovation capability in year t of enterprise j in industry i. Its calculation is as follows: Let apply0 denote the number of green patent applications in year t, then $\text{Innovation}_{ijt}$ measured by green patent applications equals $\text{Innovation}_{ijt} = \ln(\text{apply0} + 1)$. The invention patent count (invention) and the utility model patent count (utility), distinguished by patent connotation, are calculated according to the same formula.

(2) Main explanatory variable.

The main explanatory variable $\text{digital}_{jt}$ represents the degree of digital transformation of enterprise j in year t. According to existing research, the lexicon of digital transformation feature terms can be divided into two categories, namely, the underlying technology architecture and digital technology application. The underlying technology architecture refers to the ABCD digital technologies, that is, artificial intelligence, blockchain, cloud computing, and big data. The iteration and development of these technologies drive rapid change in overall digital technology and the fast development of the digital economy. These are the four most important technological cornerstones under the digital revolution, and they mainly appear in digital transformation of the enterprise back end such as research and development design, management mode, or business logic. Digital technology application refers to integration based on ABCD technologies into the market environment in which the enterprise operates, involving digital transformation and upgrading of the main business or the expansion of new digital lines of business, which mainly appears in digital transformation of front-end activities such as production and sales.

Following Wu Fei and coauthors, this paper uses text mining to compute the frequencies of digital transformation keywords in enterprise annual reports, and measures the degree of digital transformation by adding one to the frequency and then taking the natural logarithm. Since keywords in annual reports can reflect strategic features and managerial philosophies of enterprises, a higher keyword frequency in annual reports indicates greater attention and resource investment by the enterprise in that dimension, and a higher level of digital technology application.

(3) Control variables.

The vector X contains a series of other variables that may affect enterprise innovation, including enterprise size (size), enterprise age (age), board size (bm), state-ownership (state), asset-liability ratio (leverage), capital intensity (intensity), Tobin Q (tobinq), and a dummy for passing ISO 14001 (PassISO14001). The specific definitions of the control variables are provided in Table 1. In addition, industry fixed effects and year fixed effects are controlled for in Model (1), in order to control for the influence of unobservable factors that do not vary with time or industry development.

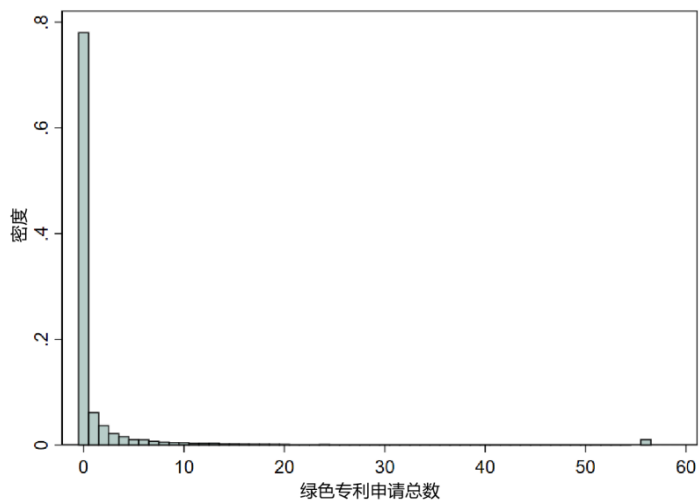

**Figure 1. Zero-inflated distribution of the number of green patent applications.**

(4) Econometric method.

Figure 1 presents the histogram of green patent applications. It is noteworthy that green patent applications display a highly skewed distribution with a large number of zeros. When the dependent variable is a count variable, the combination of log(y+1) and ordinary least squares may produce estimates that are not unbiased and that lack economic interpretability, so pseudo-Poisson regression estimation is an appropriate choice. This paper employs the Poisson Pseudo Maximum Likelihood estimator with high-dimensional fixed effects (PPMLHDFE) for empirical tests. Compared with the widely used PPML estimator for zero-inflated trade data, the high-dimensional fixed-effects Poisson estimator can test pseudo-maximum likelihood estimation more robustly. All empirical analyses in this paper are conducted using StataMP 17.0.

# IV. Empirical Results

## 4.1 Descriptive Statistics

We conduct descriptive statistics for the main variables, and the results are reported in Table 2. The mean number of green patent applications is 3.47, indicating that, during the sample period, domestic listed enterprises on average file 3.47 green patent applications per year. The standard deviation is 36.37, which far exceeds the mean, indicating substantial dispersion in green patent applications across enterprises. The median equals 0, indicating that more than half of enterprises have zero green patent applications … The coefficient of variation for the logarithm of the digital transformation keyword frequency after adding one is approximately 0.5, which indicates relatively high volatility for this variable. The mean of the state owned enterprise dummy is 0.56, indicating that 56 percent of the sample are state owned enterprises and 44 percent are private enterprises.

Table 2. Descriptive statistics of the main variables

| Variable | N | Mean | SD | Min | Median | Max |
| --- | --- | --- | --- | --- | --- | --- |
| apply0 | 15130 | 3.47 | 36.47 | 0.00 | 0.00 | 1543.00 |
| invention | 15130 | 2.35 | 29.60 | 0.00 | 0.00 | 1376.00 |
| utility | 15130 | 0.20 | 0.60 | 0.00 | 0.00 | 2.00 |
| lndigital | 15119 | 2.32 | 1.20 | 0.00 | 2.20 | 6.63 |
| lnsize | 15118 | 22.38 | 1.48 | 13.08 | 22.29 | 28.64 |
| age | 15119 | 16.05 | 5.44 | 3.00 | 16.00 | 30.00 |
| bm | 15119 | 9.73 | 2.54 | 3.00 | 9.00 | 31.00 |
| state | 15130 | 0.56 | 0.50 | 0.00 | 1.00 | 1.00 |
| leverage | 15119 | 0.46 | 0.24 | 0.00 | 0.48 | 1.76 |
| intensity | 15130 | 0.67 | 0.63 | 0.00 | 0.53 | 11.60 |
| tobinq | 15130 | 2.09 | 6.91 | 0.00 | 1.47 | 715.94 |
| PassISO14001 | 15117 | 0.18 | 0.39 | 0.00 | 0.00 | 1.00 |
| lnrd | 9828 | 17.70 | 1.91 | 5.09 | 17.86 | 25.03 |

Enterprises are sorted in ascending order by the degree of digital transformation, the sample is divided into fifteen groups, and the average number of green patent applications is calculated for each group. Figure 2 shows that the degree of digital transformation is positively correlated with the number of green patent applications, that is, the higher the degree of digital transformation, the stronger the green innovation output capability of the enterprise, which is consistent with the hypothesis stated earlier.

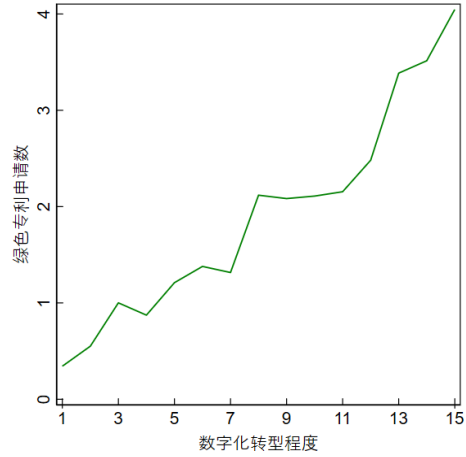

**Figure 2. Degree of digital transformation and number of green patent applications**

4.2 Baseline Regression Results

In the baseline regressions, enterprise characteristic variables are sequentially added, and Table 3 reports estimation results based on Model (1). Columns (1) through (4) separately control for characteristic variables at the levels of basic enterprise attributes, internal control, and financial condition in order to examine the linear effect of digital transformation on green innovation output capability. Column (5) reports the regression for the full specification.

According to the regression results, regardless of whether basic attributes, internal control, or financial condition are controlled, the coefficient of the main explanatory variable is significantly positive in statistical terms, the regression results are highly reliable and robust, and a positive linear relationship is present. For the regression with basic attributes (column 2), enterprise size and state ownership have significant promoting effects on green innovation output, whereas enterprise age contributes negatively. For the regression with internal-control variables (column 3), board size and having passed environmental certification significantly promote green innovation output. For the regression with financial variables (column 4), financial leverage and capital intensity significantly promote green innovation output. For the full regression (column 5), enterprise size, state ownership, board size, environmental certification, and capital intensity all have significant positive effects on green innovation output, and enterprise age is significantly negatively related to green innovation output.

To further examine whether the driving effect of digital transformation on green innovation output has temporal persistence, dynamic effects are evaluated. Forward values of the dependent variable with one to two leads are introduced, and results are presented in Table 4. The coefficients of the main explanatory variable are all positive at the 1 percent significance level, which indicates that the positive impact of digital transformation on green innovation output exists in the current year and persists for the next three years. In terms of magnitude, 0.357 is greater than 0.341, which is greater than 0.313, which is greater than 0.304. The effect is smallest in the current year, reaches the maximum in the following year, and then gradually weakens over subsequent periods. The results remain consistent when the set of control variables is not included.

Specifically, the promoting effect of digital transformation on green innovation output is at a relatively low level in the current year, reaches the maximum in the following year, and then decreases year by year. This can be explained by the time required for the transformation from research and development input to realized innovation output. In the initial stage of transformation, enterprises need to invest resources and effort to adapt to the new digital environment, adjust business processes, and train employees to master new tools and skills, so the effect in the current year may be limited. As enterprises gradually adapt and accumulate experience, green innovation capability improves, and digital technologies and data analysis are better used to improve products, processes, and services, which raises resource efficiency and protects the environment, so the effect peaks in the following year. In subsequent periods, after enterprises reach a certain level of green innovation, further improvement becomes more difficult or is constrained by other factors, so the impact gradually weakens.

Table 3. Baseline regressions

|  | (1) Baseline | (2) Basic attributes | (3) Internal control | (4) Financial condition | (5) Full |
|---|---|---|---|---|---|
| lndigital | 0.423*** | 0.320*** | 0.415*** | 0.422*** | 0.304*** |
|  | (0.0252) | (0.0248) | (0.0241) | (0.0247) | (0.0237) |
| lnsize |  | 0.736*** |  |  | 0.734*** |
|  |  | (0.0198) |  |  | (0.0222) |
| age |  | -0.0638*** |  |  | -0.0660*** |
|  |  | (0.00644) |  |  | (0.00622) |
| state |  | 0.213** |  |  | 0.176** |
|  |  | (0.0663) |  |  | (0.0638) |
| bm |  |  | 0.101*** |  | 0.0298** |
|  |  |  | (0.0103) |  | (0.00994) |
| PassISO14001 |  |  | 0.580*** |  | 0.383*** |
|  |  |  | (0.0630) |  | (0.0552) |
| leverage |  |  |  | 1.863*** | 0.131 |
|  |  |  |  | (0.117) | (0.143) |
| intensity |  |  |  | 0.477*** | 0.420*** |
|  |  |  |  | (0.0480) | (0.0448) |
| tobinq |  |  |  | -0.278*** | 0.0791** |
|  |  |  |  | (0.0329) | (0.0260) |
| _cons | 0.391*** | -15.63*** | -0.796*** | -0.374*** | -16.44*** |
|  | (0.114) | (0.443) | (0.118) | (0.0905) | (0.480) |
| Year | YES | YES | YES | YES | YES |
| Industry | YES | YES | YES | YES | YES |
| N | 14973 | 14973 | 14971 | 14973 | 14971 |
| Wald | 280.63 | 2362.16 | 465.98 | 676.64 | 2762.46 |
| p | 0.00 | 0.00 | 0.00 | 0.00 | 0.00 |

Note: *, **, and *** denote significance at the 10, 5, and 1 percent levels, respectively; the same notation applies hereafter.

Although diminishing marginal effects constrain the dynamic impact, the positive promoting effect remains strong over the long term. Enterprises should maintain patience and a long-term perspective, recognize the time lag of innovation output, set reasonable expectations for early input, and make long-term plans and preparations for green innovation investment during digital transformation. Although digital transformation requires certain investment and time costs, after the transformation enterprises can obtain more efficient innovation capability and a more competitive position, which exerts persistent positive effects and helps enterprises conduct green innovation and achieve sustainable development.

Table 4. Persistence of the effect of digital transformation on green patent applications

|  | apply0 | | apply1 | | apply2 | | apply3 | |
|---|---|---|---|---|---|---|---|---|
|  | (1) | (2) | (3) | (4) | (5) | (6) | (7) | (8) |
| lndigital | 0.423*** | 0.304*** | 0.517*** | 0.357*** | 0.497*** | 0.341*** | 0.492*** | 0.313*** |
|  | (0.0252) | (0.0237) | (0.0586) | (0.0548) | (0.0527) | (0.0461) | (0.0647) | (0.0568) |
| lnsize |  | 0.734*** |  | 1.107*** |  | 1.086*** |  | 1.029*** |
|  |  | (0.0222) |  | (0.0515) |  | (0.0536) |  | (0.0507) |
| age |  | -0.0660*** |  | -0.0227 |  | -0.0216 |  | -0.0237 |
|  |  | (0.00622) |  | (0.0122) |  | (0.0132) |  | (0.0136) |
| bm |  | 0.0298** |  | 0.0353 |  | 0.0496** |  | 0.0495** |
|  |  | (0.00994) |  | (0.0219) |  | (0.0180) |  | (0.0178) |
| state |  | 0.176** |  | 0.0978 |  | 0.0974 |  | 0.0771 |
|  |  | (0.0638) |  | (0.128) |  | (0.132) |  | (0.141) |
| leverage |  | 0.131 |  | -0.303 |  | -0.167 |  | -0.207 |
|  |  | (0.143) |  | (0.251) |  | (0.298) |  | (0.284) |
| intensity |  | 0.420*** |  | 0.980*** |  | 0.926*** |  | 0.861*** |
|  |  | (0.0448) |  | (0.0862) |  | (0.0909) |  | (0.104) |
| tobinq |  | 0.0791** |  | 0.199*** |  | 0.213*** |  | 0.188*** |
|  |  | (0.0260) |  | (0.0437) |  | (0.0480) |  | (0.0448) |
| PassISO14001 |  | 0.383*** |  | 0.240* |  | 0.302* |  | 0.335* |
|  |  | (0.0552) |  | (0.119) |  | (0.134) |  | (0.147) |
| _cons | 0.391*** | -16.44*** | 1.378*** | -25.78*** | 1.475*** | -25.36*** | 1.551*** | -23.76*** |
|  | (0.0465) | (0.480) | (0.111) | (1.300) | (0.108) | (1.412) | (0.110) | (1.378) |
| Year | YES | YES | YES | YES | YES | YES | YES | YES |
| Inudstry | YES | YES | YES | YES | YES | YES | YES | YES |
| N | 14973 | 14971 | 13487 | 13485 | 11991 | 11989 | 10493 | 12004 |
| Wald | 280.63 | 2762.46 | 77.70 | 1135.85 | 88.94 | 1033.12 | 57.67 | 809.30 |
| p | 0.00 | 0.00 | 0.00 | 0.00 | 0.00 | 0.00 | 0.00 | 0.00 |

4.3 Robustness Checks

To ensure reliability, a series of robustness checks are conducted. (1) Replacement of the dependent variable: different patent types have different requirements regarding innovation factors and environments, which leads to heterogeneous effects of digital transformation on green innovation activities. The total number of green patent applications is decomposed into the numbers of green invention patent applications and green utility-model patent applications. Considering the gestation period from input to output, one- and two-period leads are also used. As shown in columns (1) and (3) of Table 5, the degree of digital transformation significantly and positively affects the output of green invention patents over the next two years, whereas the effect on green utility-model patents is not significant.

Table 5. Robustness checks: replacing the dependent variable and sample screening

|  | ①Replace explained variables | | | | ② Manufacturing industry | ③ Exclude information and communication industry |
|---|---|---|---|---|---|---|
|  | (1) | (2) | (3) | (4) | (5) | (6) |
|  | invention1 | utility1 | Invention2 | utility2 | apply0 | apply0 |
| lndigital | 0.446*** | 0.00150 | 0.428*** | 0.000614 | 0.328*** | 0.244*** |
|  | (0.0593) | (0.00525) | (0.0510) | (0.00344) | (0.0254) | (0.0297) |
| lnsize | 1.166*** | 0.000842 | 1.144*** | -0.0324*** | 0.696*** | 0.760*** |
|  | (0.0561) | (0.00924) | (0.0578) | (0.00630) | (0.0245) | (0.0251) |
| age | -0.0180 | -0.000271 | -0.0170 | 0.00535*** | -0.0730*** | -0.0649*** |
|  | (0.0130) | (0.00166) | (0.0139) | (0.00104) | (0.00698) | (0.00719) |
| bm | 0.0584** | 0.000751 | 0.0723*** | -0.000967 | 0.0340** | 0.0358** |
|  | (0.0218) | (0.00336) | (0.0185) | (0.00207) | (0.0117) | (0.0117) |
| state | 0.243 | -0.00227 | 0.253 | -0.0121* | 0.201** | 0.0744 |
|  | (0.136) | (0.00637) | (0.139) | (0.00471) | (0.0677) | (0.0738) |
| leverage | -0.396 | -0.00893 | -0.251 | 0.0788*** | 0.560*** | 0.157 |
|  | (0.273) | (0.0298) | (0.326) | (0.0204) | (0.154) | (0.153) |
| intensity | 1.172*** | 0.00336 | 1.095*** | 0.00187 | 0.395*** | 0.405*** |
|  | (0.0920) | (0.00741) | (0.0960) | (0.00595) | (0.0505) | (0.0463) |
| tobinq | 0.227*** | -0.00682* | 0.246*** | 0.00150 | -0.00157 | 0.00869 |
|  | (0.0467) | (0.00275) | (0.0501) | (0.00127) | (0.0284) | (0.0298) |
| IsPassISO14001 | 0.246 | -0.00715 | 0.316* | 0.0208 | 0.299*** | 0.234*** |
|  | (0.134) | (0.0183) | (0.147) | (0.0123) | (0.0580) | (0.0610) |
| _cons | -28.30*** | 0.583*** | -27.84*** | 1.291*** | -15.42*** | -16.91*** |
|  | (1.379) | (0.174) | (1.473) | (0.121) | (0.522) | (0.544) |
| Year | YES | YES | YES | YES | YES | YES |
| Inudstry | YES | YES | YES | YES | YES | YES |
| N | 13485 | 3026 | 11989 | 1512 | 8748 | 13345 |
| Wald | 1164.57 | 11.50 | 1054.39 | 37.44 | 2686.08 | 2018.60 |
| p | 0.00 | 0.24 | 0.00 | 0.00 | 0.00 | 0.00 |

Green invention patents are defined as patents that are novel, useful, and environmentally friendly and generally require a higher technological level and innovation capability. Green utility-model patents are patents that solve environmental problems through practical technical solutions and improvements. Therefore digital transformation exerts a stronger promoting effect on green invention patent output that requires a higher innovation threshold, while the effect on green utility-model patent output that requires a lower innovation threshold is not significant.

(2) Retaining manufacturing enterprises: based on the fact that in 2021 mechanical and electrical

products accounted for 59 percent of China exports, manufacturing has become the most important component in the going-global industries; therefore only manufacturing enterprises are retained in column (5), and the results remain robust.

Table 6. Robustness checks: replacing the core explanatory variable with five keyword frequencies

| ④ | (1) AI | (2) Blockchain | (3) Cloud computing | (4) Big data | (5) Digital technology application |
|---|---|---|---|---|---|
| lnai | 0.360*** | | | | |
| | (0.0360) | | | | |
| lnbc | | 0.516** | | | |
| | | (0.182) | | | |
| lncc | | | 0.323*** | | |
| | | | (0.0278) | | |
| lnbd | | | | 0.327*** | |
| | | | | (0.0356) | |
| lndt | | | | | 0.244*** |
| | | | | | (0.0270) |
| lnsize | 0.778*** | 0.795*** | 0.769*** | 0.760*** | 0.757*** |
| | (0.0210) | (0.0214) | (0.0212) | (0.0215) | (0.0225) |
| age | -0.0627*** | -0.0641*** | -0.0660*** | -0.0647*** | -0.0645*** |
| | (0.00631) | (0.00650) | (0.00623) | (0.00615) | (0.00645) |
| bm | 0.0273** | 0.0267** | 0.0280** | 0.0289** | 0.0306** |
| | (0.0101) | (0.0103) | (0.00981) | (0.00987) | (0.0104) |
| state | 0.0483 | 0.00630 | 0.0928 | 0.0940 | 0.109 |
| | (0.0643) | (0.0640) | (0.0618) | (0.0639) | (0.0663) |
| leverage | 0.0734 | 0.0776 | 0.0650 | 0.0919 | 0.135 |
| | (0.144) | (0.144) | (0.143) | (0.144) | (0.142) |
| intensity | 0.423*** | 0.418*** | 0.420*** | 0.409*** | 0.412*** |
| | (0.0443) | (0.0451) | (0.0438) | (0.0443) | (0.0456) |
| tobinq | 0.0957*** | 0.104*** | 0.0985*** | 0.0987*** | 0.0867*** |
| | (0.0253) | (0.0249) | (0.0253) | (0.0253) | (0.0260) |
| IsPassISO14001 | 0.410*** | 0.421*** | 0.375*** | 0.406*** | 0.401*** |
| | (0.0554) | (0.0570) | (0.0561) | (0.0561) | (0.0561) |
| _cons | -17.12*** | -17.40*** | -16.95*** | -16.79*** | -16.79*** |
| | (0.471) | (0.480) | (0.465) | (0.480) | (0.494) |
| Year | YES | YES | YES | YES | YES |
| Inudstry | YES | YES | YES | YES | YES |
| N | 14971 | 14971 | 14971 | 14971 | 14971 |
| Wald | 2573.85 | 2239.11 | 2619.45 | 2525.50 | 2507.92 |
| p | 0.00 | 0.00 | 0.00 | 0.00 | 0.00 |

(3) Excluding information and communications enterprises: given that information and communications is a high value-added high-technology industry with intrinsically higher overseas income, enterprises in this industry are excluded in column (6), and the results remain basically unchanged.

(4) Replacement of the core explanatory variable: logarithms of keyword frequencies along five dimensions are used as alternative measures, namely Artificial Intelligence, Blockchain, Cloud Computing, Big Data, and Digital-Technology Application, and the results are highly robust.

4.4 Endogeneity Tests

Table 7. Endogeneity tests

|  | (1) lndigital | IV method (2) apply0 | LIMLmethod (3) apply0 |
|---|---|---|---|
| lnstu1 | 0.0619*** | | |
|  | (4.95) | | |
| lndigital | | 9.866*** | 9.866*** |
|  | | (4.70) | (4.70) |
| lnsize | 0.160*** | 0.154 | 0.154 |
|  | (21.18) | (0.44) | (0.44) |
| age | -0.00297 | -0.0193 | -0.0193 |
|  | (-1.61) | (-0.89) | (-0.89) |
| bm | -0.00496 | 0.130** | 0.130** |
|  | (-1.54) | (3.05) | (3.05) |
| state | -0.243*** | 2.548*** | 2.548*** |
|  | (-13.93) | (4.56) | (4.56) |
| leverage | -0.183*** | 0.313 | 0.313 |
|  | (-4.60) | (0.53) | (0.53) |
| intensity | 0.131*** | -0.650 | -0.650 |
|  | (7.31) | (-1.79) | (-1.79) |
| tobinq | 0.0384*** | -0.0725 | -0.0725 |
|  | (6.59) | (-0.70) | (-0.70) |
| IsPassISO14001 | 0.0291 | 0.765** | 0.765** |
|  | (1.37) | (2.85) | (2.85) |
| _cons | -3.410*** | -6.752 | -6.752 |
|  | (-19.08) | (-0.97) | (-0.97) |
| Year | YES | YES | YES |
| Industry | YES | YES | YES |
| Kleibergen-Paap rk LM statistic | | 26.347(0.00) | |
| Cragg-Donald Wald F statistic | | 24.482(16.38) | |
| N | 15117 | 15117 | 15117 |

Note: The value in parentheses after the Kleibergen–Paap rk LM statistic is the p value of the underidentification test, and the value in parentheses after the Cragg–Donald Wald F statistic is the ten percent critical value for the Stock–Yogo weak identification test.

Endogeneity may arise from measurement error, omitted variables, and bidirectional causality. The model includes a series of control variables and employs a two-way fixed effects specification, which reduces endogeneity caused by omitted variables. Considering the possible endogeneity from reverse causality, the number of in-school university students in the province where the enterprise is located, lagged by one period, is used as an instrumental variable. Instrumental variables estimation is first conducted. Column (1) of Table 7 shows that the instrumental variable is significantly and positively correlated with the endogenous regressor. According to column (2), the instrumental variables regression passes the underidentification test and the weak-instrument test, which indicates that the selected instrumental variable is valid. In column (2), the coefficient of digital transformation remains significantly positive, which indicates that endogeneity does not distort the conclusions. Limited information maximum likelihood estimation, which is less sensitive to weak instruments, is further used, as shown in column (3) of Table 7. The LIML estimate is consistent with the IV estimate, which indicates that digital transformation significantly promotes green innovation output and that the research findings are robust.

### 4.5 Heterogeneity Analysis

(1) Heterogeneity analysis based on temporal ordering. To enlarge intergroup differences, two subsamples covering 2010–2014 and 2015–2019 are extracted by year for regression analysis. The results show that, consistent with the baseline regressions in Table 3, the positive linear relationship between digital transformation and enterprise green innovation output is significant in both subsamples, and this relationship exhibits different trends over time. Specifically, columns (1) and (4) of Table 8 show that in the later period the estimated coefficient of digital transformation on the total number of green patent applications is smaller than that in the earlier period, which indicates that the positive effect of digital transformation on the total number of green patent applications weakens over time. In contrast, columns (2) and (5) indicate that the effect of digital transformation on the number of green invention patent applications increases over time, that is, the promoting effect of digital transformation on the number of green patent applications gradually strengthens in the later period. To further demonstrate that the promoting effect of enterprise digital transformation on its green invention patent output strengthens over time, and to eliminate the influence of level effects, the natural logarithm of the ratio of enterprise green invention patent applications (apply0) to total invention patent applications (invention) is used as the dependent variable (calculated as $\ln(\frac{invention}{apply0} + 1)$). The regression results in columns (3) and (6) of Table 8 show that the promoting effect of digital transformation on the share of green patent applications strengthens in the later period, that is, with the continuous advancement of digital transformation, the share of enterprise green invention patent output has experienced faster expansion.

This trend may be due to the continuous development of technologies and methods involved in digital transformation, as well as the accumulation of experience and knowledge by enterprises during the digital transformation process. Such experience and knowledge may help enterprises better identify and exploit opportunities related to environmental protection and, by developing new green technologies and products, enhance their capability for green invention patent output.

Table 8 Heterogeneity analysis: green innovation capability by temporal ordering

|  | Year 2010-2014 | | | Year 2015-2019 | | |
|---|---|---|---|---|---|---|
|  | (1) | (2) | (3) | (4) | (5) | (6) |
|  | apply0 | invention | inv_ratio | apply0 | invention | inv_ratio |
| *lndigital1* | 0.325*** | 0.391*** | 0.0629*** | 0.302*** | 0.497*** | 0.0762*** |
|  | (0.0428) | (0.0670) | (0.0151) | (0.0280) | (0.0682) | (0.0129) |
| *lnsize* | 0.792*** | 1.085*** | 0.0769*** | 0.704*** | 1.229*** | 0.161*** |
|  | (0.0346) | (0.0490) | (0.0160) | (0.0282) | (0.0767) | (0.0166) |
| *age* | -0.0663*** | -0.0356** | 0.00109 | -0.0660*** | -0.0143 | -0.00312 |
|  | (0.00894) | (0.0127) | (0.00448) | (0.00823) | (0.0153) | (0.00366) |
| *bm* | 0.0410** | 0.127*** | 0.0295*** | 0.0248* | 0.0441 | 0.00116 |
|  | (0.0157) | (0.0174) | (0.00709) | (0.0126) | (0.0235) | (0.00618) |
| *state* | -0.0275 | -0.388* | 0.0836 | 0.283*** | 0.378* | 0.187*** |
|  | (0.100) | (0.166) | (0.0429) | (0.0807) | (0.158) | (0.0374) |
| *leverage* | -0.313 | -0.572* | -0.0199 | 0.420* | -0.523 | -0.315*** |
|  | (0.202) | (0.289) | (0.101) | (0.191) | (0.300) | (0.0830) |
| *intensity* | 0.312*** | 1.126*** | 0.0337 | 0.500*** | 1.337*** | 0.143** |
|  | (0.0654) | (0.0984) | (0.0587) | (0.0625) | (0.176) | (0.0443) |
| *tobinq* | 0.142*** | 0.217*** | 0.0575** | 0.0359 | 0.173** | 0.0490*** |
|  | (0.0374) | (0.0475) | (0.0186) | (0.0329) | (0.0578) | (0.0123) |
| *PassISO14001* | 0.289** | 0.0460 | -0.0399 | 0.416*** | 0.230 | 0.0348 |
|  | (0.0879) | (0.159) | (0.0386) | (0.0715) | (0.179) | (0.0316) |
| _cons | -17.63*** | -26.30*** | -3.135*** | -15.82*** | -30.09*** | -4.794*** |
|  | (0.733) | (1.131) | (0.373) | (0.620) | (1.982) | (0.387) |
| *Year* | YES | YES | YES | YES | YES | YES |
| *Inudstry* | YES | YES | YES | YES | YES | YES |
| *N* | 7413 | 7354 | 1501 | 7477 | 7477 | 1854 |
| *Wald* | 1370.97 | 1132.69 | 77.54 | 1500.33 | 527.86 | 156.52 |
| *p* | 0.00 | 0.00 | 0.00 | 0.00 | 0.00 | 0.00 |

(2) Heterogeneity analysis based on enterprise size. To verify the impact of differences in enterprise size on the effect of digital transformation on green innovation output, the median of enterprise size is used as the cutoff to divide the sample into two groups, large enterprises and small and medium-sized enterprises, for separate regressions. The results in Table 9 show that digital transformation has a more significant positive impact on the green innovation output of small and medium-sized enterprises than on that of large enterprises. The possible reasons are as follows. On the one hand, the flexibility and adaptability of small and medium-sized enterprises enable them to respond more quickly to market demand and green innovation opportunities. Therefore, during digital transformation, small and medium-sized enterprises can adjust their business processes, technological applications, and organizational structures more rapidly to adapt to new market trends and environmental requirements. This agility enables them to seize opportunities for green innovation earlier. On the other hand, digital transformation can help enterprises improve production and operational efficiency, reduce resource waste and environmental pollution, and thus lower costs. Small and medium-sized enterprises usually have more limited resources and capital, so they pay more attention to cost control. Digital transformation provides opportunities for small and medium-sized enterprises to reduce production costs and environmental impacts, thereby creating more favorable conditions for their green innovation output. Through the application of digital technologies, small and medium-sized enterprises can improve energy use efficiency, waste treatment, and resource recycling in the production process, thereby further enhancing green innovation output.

Table 9 Heterogeneity analysis

|  | Enterprise Scale | | Industry Sector | |
| --- | --- | --- | --- | --- |
|  | (1) | (2) | (3) | (4) |
|  | Large-scale | Small and Medium-scale | Technology-intensive | Non-technology-intensive |
| lndigital1 | 0.271*** | 0.410*** | 0.290*** | 0.0566 |
|  | (9.98) | (8.68) | (10.38) | (1.26) |
| lnsize | 0.756*** | 0.810*** | 0.676*** | 0.839*** |
|  | (26.43) | (8.83) | (23.33) | (26.03) |
| age | -0.0542*** | -0.111*** | -0.0780*** | -0.0431*** |
|  | (-7.64) | (-8.24) | (-10.17) | (-4.35) |
| bm | 0.0352*** | -0.0212 | 0.0372** | 0.0203 |
|  | (3.32) | (-0.80) | (3.23) | (1.25) |
| state | 0.139 | 0.323* | 0.0728 | 0.405*** |
|  | (1.84) | (2.45) | (0.95) | (3.64) |
| leverage | 0.0994 | 0.333 | 0.748*** | -0.986*** |
|  | (0.59) | (1.33) | (3.89) | (-4.88) |
| intensity | 0.458*** | 0.181* | 0.597*** | 0.513*** |
|  | (8.99) | (2.19) | (7.80) | (9.61) |
| tobinq | 0.0627 | 0.0813* | 0.00457 | 0.165*** |
|  | (1.69) | (2.10) | (0.15) | (3.77) |
| IsPassISO14001 | 0.421*** | 0.145 | 0.318*** | 0.625*** |
|  | (6.78) | (1.28) | (4.91) | (6.95) |
| _cons | -17.13*** | -17.07*** | -14.82*** | -19.24*** |
|  | (-26.84) | (-8.73) | (-23.83) | (-26.97) |
| *Year* | YES | YES | YES | YES |
| *Inudstry* | YES | YES | YES | YES |
| N | 7408 | 7451 | 5486 | 9485 |

(3) Heterogeneity analysis based on industry category. Drawing on the study of Li Xuedong et al. (2018), industries are divided into three categories: technology-intensive, capital-intensive, and labor-intensive. Accordingly, the sample is divided into technology-intensive enterprises and non-technology-intensive enterprises for regression analysis. As shown in columns (3) and (4) of Table 9, in the technology-intensive enterprise sample the coefficient of lndigital is significantly positive at the 1 percent level, whereas in the non-technology-intensive enterprise sample the coefficient of the core explanatory variable is not significant. This indicates that digital transformation is more conducive to promoting the growth of green innovation output in technology-intensive industries. The possible reason is that enterprises with low technological content tend to produce products with low added value and therefore low profitability, making them unable to bear the high costs of digital transformation. These enterprises usually face heavy investment pressures in updating equipment, introducing new technologies, and training employees, and such investments may not immediately translate into significant green innovation output. As a result, these enterprises progress more slowly in digital transformation and cannot keep pace with technology-intensive enterprises. By contrast, technology-intensive enterprises possess clear advantages in digital transformation. Such enterprises generally devote substantial resources to research and development and innovation and possess technical expertise and innovative capability. Digital transformation provides them with more opportunities to improve products and services through advanced technologies and data analysis, increase production efficiency, reduce resource consumption, and achieve green innovation. Therefore, technology-intensive enterprises are more likely to obtain significant results in digital transformation and achieve growth in innovation performance.

4.6 Mechanism Tests

This paper finds that enterprise digital transformation influences final green innovation output through two channels, namely environmental management and research and development investment. The following presents two approaches to mechanism testing.

(1) Environmental management channel. Environmental management certification reflects the degree to which an enterprise values environmental protection and its capability for innovation in environmental management. Managers with strong environmental awareness will proactively engage in green innovation research and development, cleaner production, and the production of environmentally friendly products, so the green innovation effect of enterprise digital transformation will also be subject to heterogeneity. Accordingly, the sample is divided into two groups based on whether ISO 14001 environmental management certification has been passed, and regressions are estimated for the two groups. Columns (1) and (2) of Table 10 show that the coefficients are significantly positive at the 1 percent level in both groups, but the coefficient of digital transformation for enterprises that have passed ISO 14001 certification equals 0.413, whereas for enterprises without certification it equals 0.238, which is smaller. This indicates that the promoting effect of digital transformation on enterprise green innovation output is more pronounced among enterprises with environmental management certification. This implies that digital transformation promotes enterprise green innovation by advancing enterprise environmental management. Further regressions using the number of green invention patents as the dependent variable yield consistent results.

Digital transformation can help enterprises improve and innovate existing business processes, enhance data analysis capability, market insight, and customer experience, and thereby promote growth in enterprise output. Enterprises that have passed environmental management standard certification have usually already comprehensively optimized and standardized their internal environmental management and formed an effective environmental management mechanism. Therefore such enterprises are better able to rely on digital transformation to apply the outcomes of digital transformation in practical business activities.

Table 10 Heterogeneity analysis: whether environmental management system certification has been passed

|  | Passed ISO14001 | | Failed ISO14001 | |
|---|---|---|---|---|
|  | (1) | (3) | (2) | (4) |
|  | apply0 | invention0 | apply0 | invention0 |
| lndigital | 0.413*** | 0.623*** | 0.238*** | 0.408*** |
|  | (0.0369) | (0.0573) | (0.0327) | (0.0868) |
| lnsize | 0.693*** | 0.902*** | 0.748*** | 1.261*** |
|  | (0.0393) | (0.0678) | (0.0274) | (0.0662) |
| age | -0.0586*** | -0.0425** | -0.0710*** | 0.00270 |
|  | (0.00941) | (0.0152) | (0.00813) | (0.0169) |
| bm | 0.0481** | 0.0873*** | 0.0177 | 0.0567** |
|  | (0.0177) | (0.0247) | (0.0123) | (0.0204) |
| state | 0.402*** | 0.765*** | 0.0428 | -0.0226 |
|  | (0.0954) | (0.178) | (0.0822) | (0.161) |
| leverage | 0.211 | 0.181 | 0.119 | -0.728** |
|  | (0.237) | (0.390) | (0.176) | (0.263) |
| intensity | 0.358*** | 0.474*** | 0.442*** | 1.314*** |
|  | (0.0784) | (0.130) | (0.0535) | (0.102) |
| tobinq | 0.0175 | 0.135* | 0.0920** | 0.159** |
|  | (0.0422) | (0.0613) | (0.0330) | (0.0594) |
| _cons | -15.51*** | -22.10*** | -16.44*** | -30.59*** |
|  | (0.772) | (1.529) | (0.624) | (1.813) |
| Year | YES | YES | YES | YES |
| Inudstry | YES | YES | YES | YES |
| N | 2758 | 2758 | 12200 | 12200 |

(2) Research and development investment channel. A three-step approach is used to conduct the mediation mechanism test. On the basis of the baseline regression model, the enterprise research and development investment variable is added. If, after adding the mediator, the coefficient of the main explanatory variable decreases, this indicates that increasing research and development investment is one of the channels through which digital transformation promotes enterprise green innovation output. To test this channel, data on enterprise research and development investment are obtained, and the logarithm of research and development investment is used as the mediator variable. The regression results of the mediation effect model are shown in Table 11. Column (1) shows that the degree of enterprise digital transformation is significantly positively related to green innovation output. Column (2) shows that the degree of enterprise digital transformation is significantly positively related to research and development investment. Column (3) shows that both digital transformation and research and development investment have significantly positive coefficients, but the coefficient of digital transformation is smaller than in column (1). This indicates that the mediation effect holds. Therefore the regression results in Table 11 confirm that the mediation effect of digital transformation promoting enterprise green innovation output through increased research and development investment is valid.

The reasons why enterprise digital transformation promotes an increase in research and development investment may include the following. On the one hand, enterprise digital transformation makes enterprises pay greater attention to research and development activities, especially research and development activities in environmentally friendly and green technologies, thereby increasing investment in green technologies. On the other hand, digital transformation can improve the efficiency and precision of the technologies and methods used in the research and development process, thereby further increasing the output efficiency of research and development investment.

Table 11 Digital transformation, research and development investment, and green innovation output

|  | (1) lnrd | (2) apply0 | (3) apply0 |
|---|---|---|---|
| *lndigital* | 0.279*** | 0.304*** | 0.216*** |
|  | (0.0164) | (0.0237) | (0.0239) |
| *lnrd* |  |  | 0.514*** |
|  |  |  | (0.0362) |
| *lnsize* |  | 0.734*** | 0.241*** |
|  |  | (0.0222) | (0.0408) |
| *age* |  | -0.0660*** | -0.0514*** |
|  |  | (0.00622) | (0.00662) |
| *bm* |  | 0.0298** | 0.0208* |
|  |  | (0.00994) | (0.00975) |
| *state* |  | 0.176** | 0.244*** |
|  |  | (0.0638) | (0.0658) |
| *leverage* |  | 0.131 | 0.297 |
|  |  | (0.143) | (0.159) |
| *intensity* |  | 0.420*** | 0.158** |
|  |  | (0.0448) | (0.0557) |
| *tobinq* |  | 0.0791** | 0.00537 |
|  |  | (0.0260) | (0.0282) |
| *IsPassISO14001* |  | 0.383*** | 0.330*** |
|  |  | (0.0552) | (0.0562) |
| _cons | 14.70*** | -16.44*** | -13.27*** |
|  | (0.154) | (0.480) | (0.521) |
| Year | YES | YES | YES |
| Industry | YES | YES | YES |
| N | 9828 | 14971 | 9780 |
| $R^2$ | 0.181 |  |  |
| Wald |  | 2762.46 | 2863.23 |
| p | 0.00 | 0.00 | 0.00 |

# V. Research Conclusions and Policy Implications

Digital transformation has created opportunities for innovation and development for enterprises, driving companies to continuously reach new heights in the field of green innovation. Based on financial and annual report data from 1,512 A-share listed companies in China from 2010-2019, this paper constructs enterprise digital transformation indicators and time and industry dual fixed effects models to examine the impact and mechanisms of enterprise digital transformation on green innovation output. The research results show: (1) Enterprise digital transformation can significantly promote enterprise green innovation output, and this conclusion remains valid after a series of robustness tests; (2) The dynamic impact of digital transformation on enterprise green innovation output cannot avoid the limitation of diminishing marginal effects, but its positive promoting effect still has strong temporal continuity; (3) Mechanism testing shows that digital transformation can enhance enterprise green innovation output by increasing enterprise R&D investment and strengthening environmental management; (4) Heterogeneity testing finds that digital transformation has differential impacts on enterprise green innovation due to differences in the technological level of the industry to which the enterprise belongs and the enterprise's environmental management capabilities, specifically manifested in that digital transformation has a more obvious promoting effect on green innovation output of small and medium-sized enterprises in technology-intensive industries.

Based on this, this paper proposes the following policy recommendations for the future development of enterprise digital transformation: First, strengthen the popularization and promotion of digital technology, encourage enterprises to apply digital technology to improve R&D investment and green innovation output. Government departments can help enterprises improve their digitalization level by providing digital technology training and financial support. Second, promote environmental management certification, encourage enterprises to implement environmental management measures to reduce environmental pollution and resource waste. Governments can promote enterprise environmental management implementation through strengthening the formulation and implementation of environmental protection laws and regulations, as well as providing environmental protection incentives and other policies. Third, support the application and utilization of green patents to improve enterprises' green innovation capabilities and competitiveness. Government departments can promote enterprise green patent applications and utilization by providing exemptions for green patent application fees and policy support for green patent utilization. Fourth, enterprises should view digital transformation as a long-term strategy and formulate corresponding plans and measures to achieve sustainable green innovation and sustainable development goals. Because digital transformation is not only about responding to current market demands and competitive pressures, but more importantly, about achieving long-term sustainable development. Actively promoting digital transformation can enhance enterprises' green innovation output and gain greater competitive advantages in the future.